\documentclass[aps, prl, reprint, superscriptaddress]{revtex4-2}

\usepackage{graphicx}
\usepackage{bm}
\usepackage{amsmath}
\usepackage{amssymb}
\usepackage{xcolor}
\usepackage[normalem]{ulem}

\newcommand{\rmi}{{\rm i}}
\newcommand {\e}{{\rm e}}

\newcommand{\vBext}[0]{\mathbf{B_0}}
\newcommand{\Bext}[0]{B_x}

\newcommand{\VSi}[0]{\mathrm{V_{Si}}}

\newcommand{\ket}[1]{|#1\rangle}
\newcommand{\av}[1]{\langle #1 \rangle}

\begin{document}

\title{Dynamical Reorientation of Spin Multipoles in Silicon Carbide by Transverse Magnetic Fields}
\author{A.~Hern\'{a}ndez-M\'{i}nguez}
\email{hernandez-minguez@pdi-berlin.de}
\affiliation{Paul-Drude-Institut f\"{u}r Festk\"{o}rperelektronik, Leibniz-Institut im Forschungsverbund Berlin e.V., Hausvogteiplatz 5-7, 10117 Berlin, Germany}
\author{A.~V.~Poshakinskiy}
\affiliation{ICFO-Institut de Ciencies Fotoniques, The Barcelona Institute of Science and Technology, 08860 Castelldefels, Barcelona, Spain}
\author{M.~Hollenbach}
\affiliation{Helmholtz-Zentrum Dresden-Rossendorf, Institute of Ion Beam Physics and Materials Research, Bautzner Landstrasse 400, 01328 Dresden, Germany}
\author{P.~V.~Santos}
\affiliation{Paul-Drude-Institut f\"{u}r Festk\"{o}rperelektronik, Leibniz-Institut im Forschungsverbund Berlin e.V., Hausvogteiplatz 5-7, 10117 Berlin, Germany}
\author{G.~V.~Astakhov}
\affiliation{Helmholtz-Zentrum Dresden-Rossendorf, Institute of Ion Beam Physics and Materials Research, Bautzner Landstrasse 400, 01328 Dresden, Germany}

\date{\today}

\begin{abstract}

The long-lived and optically addressable high-spin state of the negatively charged silicon vacancy ($\VSi$) in silicon carbide makes it a promising system for applications in quantum technologies. Most studies of its spin dynamics have been performed in external magnetic fields applied along the symmetry axis. Here, we find that the application of weak magnetic fields perpendicular to the symmetry axis leads to nontrivial behavior caused by dynamical reorientation of the $\VSi$ spin multipole under optical excitation. Particularly, we observe the inversion of the quadrupole spin polarization in the excited state and appearance of the dipole spin polarization in the ground state. The latter is much higher than thermal polarization and cannot be induced solely by optical excitation. Our theoretical calculations reproduce well all sharp features in the spin resonance spectra, and shine light on the complex dynamics of spin multipoles in these kinds of solid-state systems.         

\end{abstract}

\maketitle


Atom-like defects in solids are attractive for various quantum applications, including quantum sensing, quantum communication, and potentially quantum computation. The electronic structure of the ground and the optically accessible excited states (GS and ES, respectively) of these centers is typically composed of a few electrons with a well-defined total spin $S>1/2$. Celebrated examples include the nitrogen-vacancy antisite in diamond and the silicon divacancy in silicon carbide (SiC), both with $S=1$, as well as the negatively charged silicon vacancy ($\VSi$) in SiC with $S=3/2$ \cite{Koehl_MB40_1146_15,10.1038/s41578-018-0008-9, 10.1038/s41566-018-0232-2e8e, Zhang_APR7_31308_20}. A prominent property of these centers is the ability to initialize spin states by excitation with unpolarized light~\cite{10.1103/physrevb.83.125203, 10.1038/nature10562, 10.1103/physrevlett.109.226402}. Microscopically, the optically induced spin initialization is caused by spin-dependent nonradiative relaxation from the ES to the GS via metastable states (MSs), and its orientation is settled by the defect symmetry axis.

Importantly, the optical excitation does not generate the conventional dipole spin polarization, but rather the quadrupole spin polarization, i.e. the alignment of the defect spin along the symmetry axis~\cite{10.1038/s41467-019-09429-x}. Obviously, if the external magnetic field is also applied along the symmetry axis, as in most optically detected magnetic resonance (ODMR) experiments on the $\VSi$ center~\cite{10.1038/ncomms8578, 10.1103/physrevb.93.081207, 10.1103/physrevb.99.184102, 10.48550/arxiv.2307.13648}, the sign of the quadrupole spin polarization does not change. However, the application of a transverse magnetic field induces the reorientation of the spin quadrupole, which occurs at different field strengths for the GS and ES. Recent experiments using transverse magnetic fields~\cite{10.1103/physrevlett.125.107702, 10.1126/sciadv.abj5030, Vasselon2023} suggest that this reorientation, combined with the dynamical coupling of the GS and ES spin via the optical excitation, can result in an unusual behavior of the measured ODMR spectra, like the observation of negative spin resonances in the ES~\cite{10.1126/sciadv.abj5030, Vasselon2023}. In this contribution, we experimentally observe and theoretically explain the dynamical reorientation of the quadrupole spin polarization at weak transverse magnetic fields, which manifests itself as an inversion of the ODMR contrast in the ES. These results could contribute to a better understanding of the different behaviors observed for the ES spin resonances~\cite{10.1103/physrevx.6.031014, 10.1063/5.0027603}. Furthermore, we detect a non-zero dipole spin polarization, which is forbidden under linearly polarized excitation and is orders of magnitude stronger than the thermally induced polarization due to the Zeeman splitting. 

To demonstrate these effects, we investigate the spin dynamics of the $\VSi$ defect in 4H-SiC with cubic local crystallographic environment (also known as the V2 center)~\cite{Ivady:2017bq}. The GS of the V2 center has been proposed as quantum magnetometer~\cite{10.1038/srep05303} with significantly improved performance by optimizing the material properties~\cite{10.1103/physrevapplied.15.064022, 10.1103/physrevapplied.19.044086}, and used in different modalities, including vector magnetometry~\cite{10.1103/physrevapplied.4.014009, 10.1103/physrevapplied.6.034001}, microwave-free magnetometry~\cite{10.1103/physrevx.6.031014} and fiber-integrated magnetometry~\cite{10.1364/ol.476305}. In contrast to the GS, the ES of the V2 center has received much less attention, although its huge thermal shift~\cite{10.1038/srep33301} makes it ideal for quantum thermometry~\cite{10.1038/srep05303, 10.1063/5.0027603}, and the ES is much more sensitive to strain than the GS~\cite{10.1063/5.0040936, 10.1126/sciadv.abj5030}. Furthermore, the joint spin dynamics of the GS and ES under simultaneous excitation by optical and microwave fields is crucial for the realization of spin-photon interfaces~\cite{10.1038/s41467-019-09873-9, 10.1038/s41534-022-00534-2}.

\begin{figure}
\includegraphics[width=\linewidth]{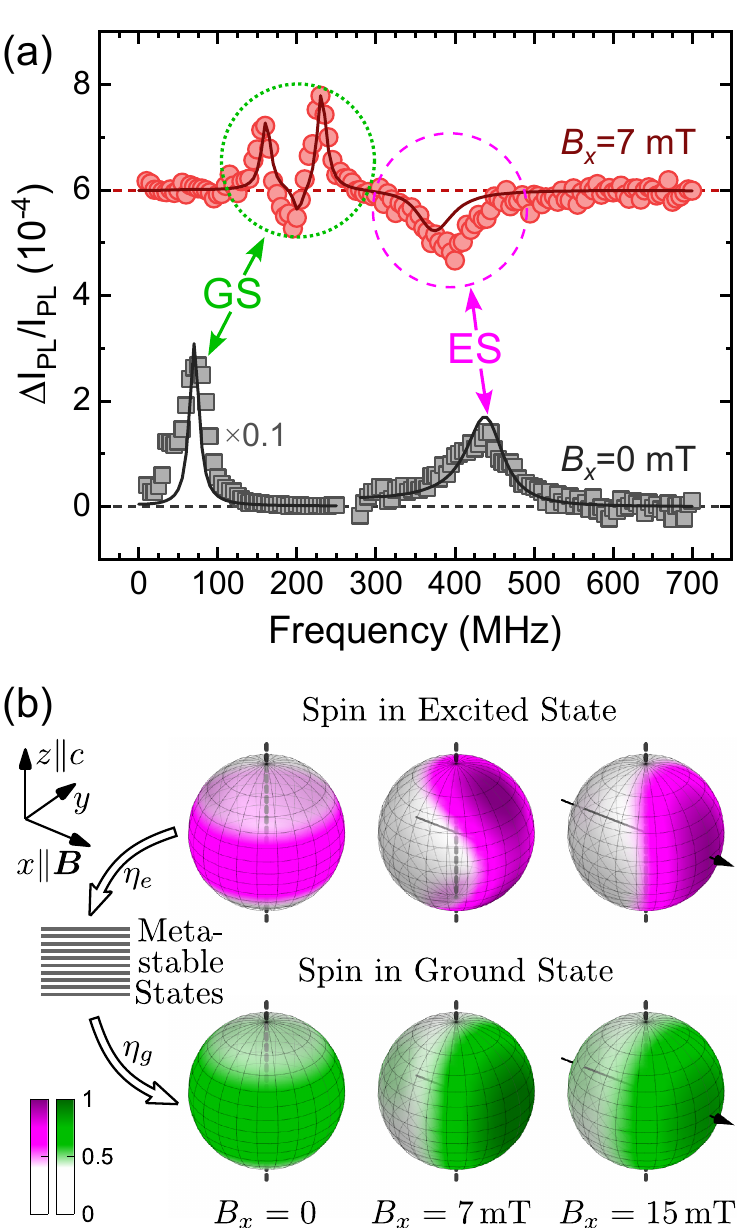}
\caption{(a) ODMR spectra as a function of the MW frequency in the absence (black squares) and under a magnetic field, $\Bext$, transverse to the $c$-axis of the 4H-SiC (red circles). The solid curves are the ODMR spectra obtained from the theoretical model. The data are vertically shifted for clarity. (b) Husimi maps of the GS and ES spin distributions (green and magenta spheres, respectively), for $\Bext=0$~mT, 7~mT and 15~mT, generated by optical excitation and spin-selective relaxation via the metastable states with efficiencies $\eta_g$ and $\eta_e$. The vertical dashed lines indicate the $\hat{\bm z}$ direction, and the horizontal arrows denote the direction and strength of $\Bext$.}
\label{fig1}
\end{figure}

Figure~\ref{fig1} summarizes the effect of applying an external magnetic field $\vBext=(\Bext,0,0)$ perpendicular to the $c$-axis of the 4H-SiC ($\hat{\bm z}$ is chosen parallel to the $c$-axis). In the absence of $\Bext$, the ODMR spectrum (see Supplemental Material (SM) for a description of the experimental setup~\footnote{See Supplemental Material at URL for details about experimental setup, Husimi spin distributions, kinetic equations for the spin density matrices, rate equations for the spin-level populations, and extraction of spin dipole and quadrupole from the intensities of the ODMR signal.}) consists of two positive resonances centered at 70~MHz and 440~MHz, see black squares in Fig.~\ref{fig1}(a). These resonances correspond to the spin transitions between the $\ket{\pm1/2}_z$ and $\ket{\pm3/2}_z$ spin doublets in the GS and ES~\cite{10.1103/physrevx.6.031014}. As soon as $\Bext$ is applied (red circles in Fig.~\ref{fig1}(a)), the ES resonance changes sign, and the GS resonance splits into a doublet with an additional, negative resonance between the two positive peaks, see regions marked with magenta dashed and green dotted circles, respectively.

The differences in the two ODMR spectra of Fig.~\ref{fig1} are a consequence of the reorientation of the GS and ES spin multipoles under the transverse magnetic field. This is illustrated in Fig.~\ref{fig1}(b), which shows the distributions of spin quasiprobabilities on the surface of a sphere following the Husimi representation~\cite{Husimi1940} (see SM for the calculation of the Husimi spin distributions~\cite{Note1}). The north and south poles correspond to the $\ket{+3/2}_z$ and $\ket{-3/2}_z$ states, respectively, and the equator represents the $\ket{\pm 1/2}_z$ states. In the absence of $\Bext$, the optical excitation and the spin-selective relaxation via the MSs lead to a steady-state spin distribution with the $\ket{\pm 1/2}_z$ spin doublets preferentially populated for both GS and ES, see the spheres in the first column of Fig.~\ref{fig1}(b). This distribution corresponds to a zero dipole spin polarization (the Husimi representation is symmetric with respect to all three $xy$, $yz$ and $xz$ planes), but to a non-zero quadrupole spin polarization (different spin densities at the poles and the equator).

For $\Bext=7$~mT, the reorientation of the spin toward $\hat{\bm x}$ generates an asymmetric spin distribution with respect to the $yz$ plane for both GS and ES, see the spheres in the second column of Fig.~\ref{fig1}(b), and thus a non-zero dipole spin polarization along $\hat{\bm x}$. More importantly, while the maximum of the GS spin distribution remains on the equator of the Husimi sphere, the maximum quasiprobabilities in the ES spin are shifted toward the poles, thus indicating a quadrupole spin polarization of opposite sign to that at $\Bext=0$. At larger magnetic fields, the reorientation of the ES spin is almost fulfilled, and its Husimi distribution is again similar to that of the GS spin, see the spheres in the third column of Fig.~\ref{fig1}(b).

\begin{figure}
\includegraphics[width=\linewidth]{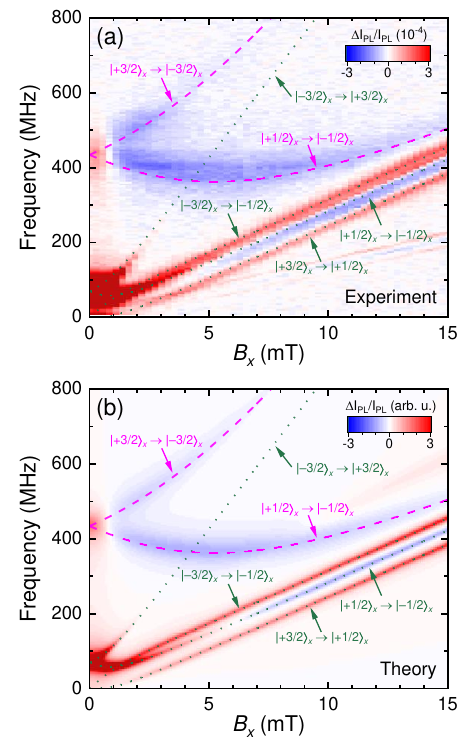}
\caption{(a) Experimental measurement of the relative change in PL intensity as a function of MW frequency and magnetic field, $B_x$. (b) Change in PL intensity predicted by the theoretical model. The green dotted and magenta dashed lines in both panels denote the magnetic field dependencies of the relevant GS and ES spin transitions estimated using Eq.~\eqref{eq:Hamil}.}
\label{fig2}
\end{figure}

The ODMR signal for the full range of applied magnetic field strengths and microwave (MW) excitation frequencies is shown in the two-dimensional plot of Fig.~\ref{fig2}(a). As already seen in Fig.~\ref{fig1}(a), the ODMR spectrum for $\Bext=0$ consists of positive resonances (red color) for both the GS and ES spin transitions. The application of $\Bext$ splits these resonances into new ones, several of them with negative values of the ODMR signal (blue color). Note that, due to the large power of the applied MW, the ODMR measurements of Figs.~\ref{fig2}(a) and \ref{fig1}(a) are also sensitive to non-linear effects taking place at half the frequency of the GS resonances.

The superimposed green dashed and magenta dotted curves display the magnetic field dependencies of the GS and ES spin resonances, respectively, estimated from the spin Hamiltonian

\begin{equation}\label{eq:Hamil}
\mathcal{H}^{(g,e)}=g^{(g,e)}\mu_B S_x B_x + hD^{(g,e)}\left(S_z^2-\frac{5}{4}\right).
\end{equation}

\begin{figure}
\includegraphics[width=\linewidth]{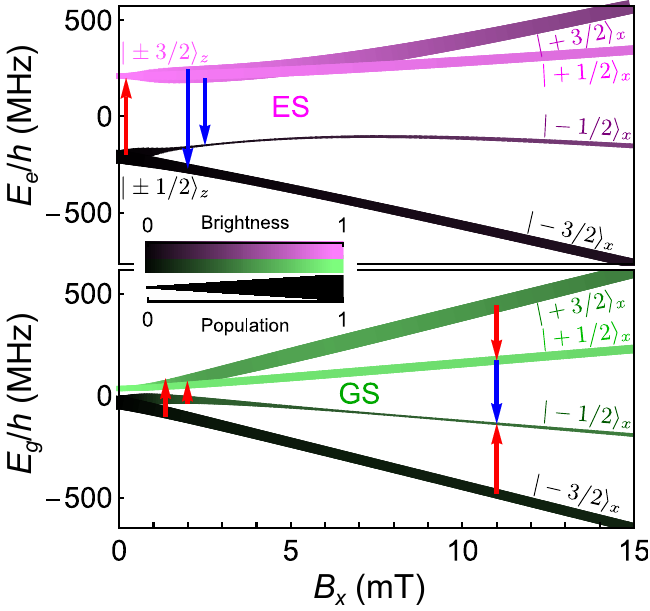}
\caption{Fine energy structure of the GS and ES spin eigenstates (lower and upper panels, respectively) under a transverse magnetic field, $\Bext$. The line thicknesses are proportional to the population densities of the corresponding spin states according to our theoretical model. The color brightnesses are proportional to the PL intensities for optical transitions between ES and GS with the same spin projection.}
\label{fig3}
\end{figure}

\noindent Here, $g^{(g,e)} \approx 2$ is the $g$-factor, $\mu_B$ the Bohr magneton, $h$ the Planck constant, and $D^{(g)}=35$~MHz, $D^{(e)}=220$~MHz are the room temperature zero-field splitting constants stemming from the crystal field for the GS and ES, respectively. The green and magenta curves in Fig.~\ref{fig3} show the energy dependence of the spin eigenstates for the GS (lower panel) and ES (upper panel), respectively, according to Eq.~\eqref{eq:Hamil}. In the absence of $\Bext$, the four spin eigenstates are split into two Kramers doublets $\ket{\pm 1/2}_z$ and $\ket{\pm 3/2}_z$ in both GS and ES. The application of $\Bext$ turns the spin quantization axis towards $\hat{\bm x}$. The substantial change in the spin structure occurs when the Zeeman term overcomes the zero-field term in Eq.~\eqref{eq:Hamil}. This happens at $\Bext > hD^{(g)}/(g\mu_B) = 1.25$~mT for the GS, and at a much larger $\Bext > hD^{(e)}/(g\mu_B) = 7.86$~mT for the ES. However, for the sake of simplicity, we will use the projection along $\hat{\bm x}$ at $\Bext \to \infty$ to label both GS and ES spin states under $\Bext\neq0$.

Taking into account the calculated GS spin eigenstates in Fig.~\ref{fig3}, and the fact that their spin quantization axis is practically $\hat{\bm x}$ at $\Bext > 1.25$~mT, the only allowed spin resonances above this magnetic field value are those fulfilling $\Delta S_x = \pm 1$, similar to the case when $\vBext$ is applied along $\hat{\bm z}$~\cite{Tarasenko_pssb255_1700258_2018}. Therefore, the positive doublet observed in Fig.~\ref{fig2}(a) corresponds to the $\ket{+3/2}_x\rightarrow\ket{+1/2}_x$ and $\ket{-3/2}_x\rightarrow\ket{-1/2}_x$ spin transitions, see the red vertical arrows in the lower panel of Fig.~\ref{fig3}. However, an additional, negative resonance  now appears at the center of the doublet, which corresponds to the $\ket{+1/2}_x\rightarrow\ket{-1/2}_x$ spin transition (blue arrow in the lower panel of Fig.~\ref{fig3}). The presence of this new resonance in the ODMR spectrum of Fig.~\ref{fig2}(a) indicates a population difference between the $\ket{+1/2}_x$ and $\ket{-1/2}_x$ spin states, and thus an optically generated dipole spin polarization.

Regarding the ES spin resonance, it changes sign at $\Bext\approx1$~mT, see transition from red to blue in Fig.~\ref{fig2}(a), and then splits into two branches. According to the spin Hamiltonian of Eq.~\eqref{eq:Hamil}, these resonances correspond to the $\ket{+3/2}_x \rightarrow \ket{-3/2}_x$ and $\ket{+1/2}_x \rightarrow \ket{-1/2}_x$ spin transitions, see blue vertical arrows in the upper panel of Fig.~\ref{fig3}. The $\ket{+3/2}_x \rightarrow \ket{-3/2}_x$ spin resonance is only allowed at low $\Bext$, where the mixing of the spin states leads to a nonzero matrix element for this transition (such resonance is also observed in the GS at $\Bext<2$~mT). Above $\Bext\approx8$~mT, this resonance is almost suppressed and only the $\ket{+1/2}_x \rightarrow \ket{-1/2}_x$ spin transition fulfilling the condition $\Delta S_x=\pm1$ is still observed.

To better understand the new features observed in Fig.~\ref{fig2}(a), we have modeled the spin dynamics of the $\VSi$ center using the spin-density matrix approach. In the model, both GS and ES are described by the $4\times 4$ spin-density matrices $\rho_g$ and $\rho_e$, corresponding to the total electron spin $3/2$. The model also includes the dark intermediate MS with spin $S=1/2$, described by a single occupation number $N_m$ \footnote{This approximation is justified if the thermalization rate within the MS is much faster than the transition rate towards the GS.}. In the absence of $\Bext$, the optical transitions between GS and ES are spin conserving, while the nonradiative relaxation via the MS is spin dependent. We therefore assume larger relaxation rates via the MS for the $\ket{\pm 1/2}_z$ states than for the $\ket{\pm 3/2}_z$ ones. Under these conditions, the spin dynamics is governed by the set of kinetic equations described in the SM~\cite{Note1}.

The ODMR spectra are calculated by including in the model a magnetic field in the form $\bm B = \bm B_0 + \bm B_1 \e^{-\rmi\omega t} + \bm B_1^* \e^{\rmi\omega t}$, where $\bm B_0\parallel \hat{\bm x}$ is the static field and $\bm B_1 \parallel \hat{\bm y}$ is the MW field that is assumed to be small. We solve the kinetic equations up to terms $\sim B_1 B_1^*$ and find the corresponding correction to the photoluminescence (PL) intensity $I_\mathrm{PL}=\Gamma\,{\rm Tr\,}\rho_e$, with $\Gamma$ being the optical recombination rate. Figure~\ref{fig2}(b) shows the results of our calculation. It reproduces all the features of Fig.~\ref{fig2}(a), including the change of sign at $\Bext\approx 1$~mT for the ES resonances and the appearance of the negative resonance in the GS. 

In addition to the energy of the spin eigenstates, we also show in Fig.~\ref{fig3} the calculated steady-state population densities of the spin sublevels (proportional to the thickness of the curves), and the intensity of the optical transitions between ES and GS (proportional to the color brightness). The MW magnetic field brings the $\VSi$ center from the more populated spin state towards the less populated one, see the directions of the red and blue vertical arrows in Fig.~\ref{fig3}. The sign of the ODMR signal is positive if the final spin state is brighter than the initial state (red arrows), and negative in the opposite case (blue arrows). Our spin model leads to a steady-state spin distribution at $\Bext=0$ with the $\ket{\pm 1/2}_z$ states preferentially populated in both the GS and ES, cf. the thicker curves for the $\ket{\pm 1/2}_z$ eigenstates in Fig.~\ref{fig3} and the spin distributions around the equator of the spheres in Fig.~\ref{fig1}(b). This spin configuration corresponds to a negative quadrupole spin polarization, defined as $\av{\delta S_z^2}\equiv \frac13 \av{2S_z^2 -S_x^2 - S_y^2} = \av{S_z^2-5/4}$, and calculated according to the formula:

\begin{equation}\label{eq:dSz2}
\av{\delta S_z^2}=\frac{n_{\ket{\pm 3/2}_z}-n_{\ket{\pm 1/2}_z}}{n_{\ket{\pm 3/2}_z}+n_{\ket{\pm 1/2}_z}}.
\end{equation}

\noindent 
Here, $n_{\ket{\pm 3/2}_z}$ and $n_{\ket{\pm 1/2}_z}$ are the corresponding spin population densities. Since the $\ket{\pm 3/2}_z$ states are brighter than the $\ket{\pm 1/2}_z$ ones, the model predicts a positive ODMR signal in both GS and ES, see red vertical arrows at $\Bext=0$ in Fig.~\ref{fig3}. Moreover, $n_{\ket{+3/2}_z}=n_{\ket{-3/2}_z}$ and $n_{\ket{+1/2}_z}=n_{\ket{-1/2}_z}$, which means that the optical excitation does not create any dipole spin polarization.

The spin distributions in the absence of $\Bext$ change significantly as soon as the transverse magnetic field is applied. The reorientation of the spin towards $\hat{\bm x}$ causes a reorganization of the population densities, with the $\ket{\pm 3/2}_x$ states being now more populated than the $\ket{\pm 1/2}_x$ states, see the thicker curves for the $\ket{\pm 3/2}_x$ eigenstates at large magnetic fields in Fig.~\ref{fig3}. In addition, $n_{\ket{+3/2}_x} \neq n_{\ket{-3/2}_x}$ and $n_{\ket{+1/2}_x} \neq n_{\ket{-1/2}_x}$. Therefore, the application of $\Bext$ generates a dipole spin polarization along $\hat{\bm x}$, which causes the appearance of the $\ket{+3/2}_x \rightarrow \ket{-3/2}_x$ and $\ket{+1/2}_x \rightarrow \ket{-1/2}_x$ spin resonances in the ODMR spectra of Fig.~\ref{fig2}(a), as long as their transition matrix elements are nonzero.

We focus now on the sign change in the ES spin resonance. According to our model, it happens at the magnetic field where the population densities of all four ES spin eigenstates are similar (same curve thickness in the upper panel of Fig.~\ref{fig3}). The value of this magnetic field is determined by the relative efficiencies of the two mechanisms involved in the formation of quadrupole spin polarization shown in Fig.~\ref{fig1}(b): (i) depletion of the ES $\ket{\pm 1/2}_z$ population with efficiency $\eta_e$ due to the spin-dependent transitions from the ES to the MS, and (ii) increment of the GS $\ket{\pm 1/2}_z$ population with efficiency $\eta_g$ caused by the spin-dependent transitions from the MS to the GS. Moreover, at $\Bext=0$, the spin-preserving optical transitions lead to an efficient transfer of the spin populations between GS and ES. By assuming that $\eta_g > \eta_e$, our model results in steady-state spin populations with $n_{\ket{\pm 1/2}_z} > n_{\ket{\pm 3/2}_z}$ for both the GS and ES at $ \Bext=0$. However, the fast reorientation of the spin quantization axis in the GS at low transverse magnetic fields suppresses the spin transfer from the GS to the ES. Under these conditions, we get $n_{\ket{\pm 3/2}_z} > n_{\ket{\pm 1/2}_z} $ in the ES, and thus a reversal of the quadrupole spin polarization defined in Eq.~\eqref{eq:dSz2}. The transition between both regimes happens at a magnetic field determined by the ratio $\eta_e/\eta_g$ according to the formula~\cite{Note1}:
 
\begin{equation}\label{eq:etaratio}
\frac{\eta_e}{\eta_g} = \frac{1}{4} \left( 1+ \frac3{1+b_g^2+b_g^4}\right),
\end{equation}

\noindent with $b_g=g\mu_B\Bext/hD^{(g)}$. Introducing the experimentally obtained values of $\Bext \approx 1$~mT and $D^{(g)} \approx 35$~MHz in Eq.~\eqref{eq:etaratio}, we obtain $\eta_e/\eta_g\approx 0.7$.
 
\begin{figure}
\includegraphics[width=\linewidth]{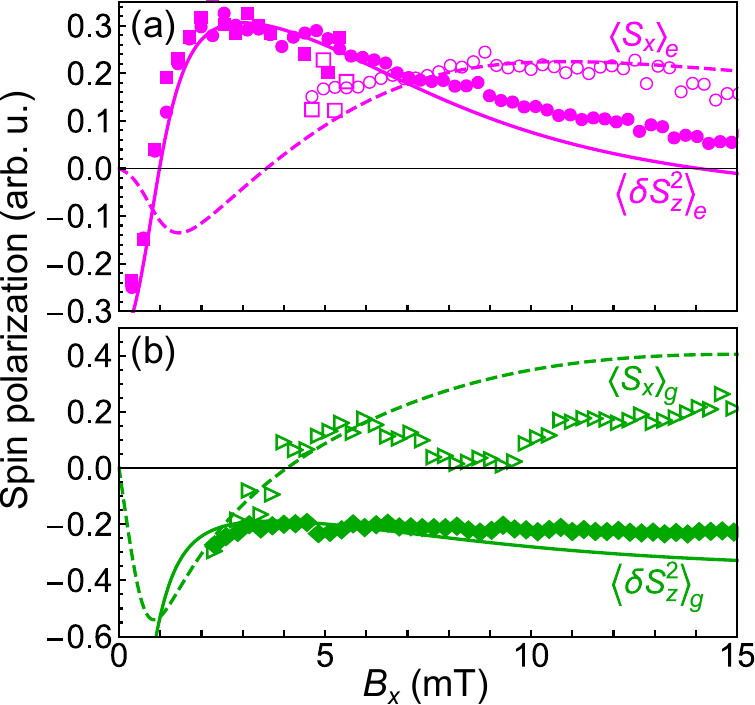}
\caption{(a) Calculated magnetic field dependencies of the steady state quadrupole spin polarization, $\av{\delta S_z^2}_e$ (solid magenta curve), and dipole spin polarization along $x$, $\av{S_x}_e$ (dashed magenta curve), in the ES. The solid and open symbols are the experimental values extracted from the measured ODMR spectra in Fig.~\ref{fig2}(a). (b) Same as (a), but for the GS spin multipoles (solid green curves and solid symbols for $\av{\delta  S_z^2 }_g$, dashed green curve and open symbols for $\av{S_x}_g$).}
\label{fig4}
\end{figure}

Finally, Figs.~\ref{fig4}(a) and \ref{fig4}(b) show the magnetic field dependencies of the out-of-plane quadrupole spin polarization (solid curves) and the in-plane dipole spin polarization (dashed curves) for the ES and GS, respectively, predicted by our spin density matrix model. The solid and open symbols show the experimental values estimated from the peak areas of the ODMR spectra (see SM for the details of the theoretical calculation and the procedure to extract the experimental values~\cite{Note1}). While the GS quadrupole spin polarization is always negative for the whole range of magnetic fields studied (green solid curve), the ES quadrupole spin polarization becomes positive at magnetic fields larger than 1~mT (magenta solid curve). This result confirms that the reorientation of the spin quadrupole polarization under the transverse magnetic field is responsible for the sign change of the ES resonances in the ODMR spectrum.

As for the dipole spin polarization, its appearance is responsible for the observation of the $\ket{+1/2}_x \rightarrow \ket{-1/2}_x$ and $\ket{+3/2}_x \rightarrow \ket{-3/2}_x$ resonances in the ODMR spectra of Fig.~\ref{fig2}(a). Moreover, our model predicts an inversion of the dipole spin polarization for both GS and ES at $\Bext \approx 4$~mT, which is responsible for the sign change of the GS $\ket{+1/2}_x \rightarrow \ket{-1/2}_x$ spin resonance predicted in Fig.~\ref{fig2}(b). Such an inversion is not clearly seen in the experimental data of Fig.~\ref{fig2}(a) due to the overlap of the spin resonances at weak magnetic fields. However, the dipole spin polarization is nonzero only over a limited magnetic field range and this resonance should vanish when $g\mu_B\Bext$ is much larger than both $hD^{(g)}$ and $hD^{(e)}$. 

In conclusion, we show that the dynamical reorientation of the spin multipoles by a magnetic field transverse to the symmetry axis of the $\VSi$ center leads to the appearance of new, sharp features in the ODMR spectra. The experimental results are well reproduced by our theoretical model, which explains the nontrivial transformation of the ODMR by the inversion of the quadrupole spin polarization and the appearance of a dipole spin polarization within certain magnetic field ranges. The understanding of the multipole spin dynamics in the GS and ES under optical excitation is important for the realization of simultaneous magnetic field and temperature sensing using the same spin center as well as for the implementation of spin-photon interfaces. Our theoretical model is not limited to the 3/2-spin of the $\VSi$, and we expect similar effects to happen for any color center with $S>1/2$ and a well defined symmetry axis.

\begin{acknowledgments}
The authors would like to thank H. Tornatzky for a critical reading of the manuscript. G.V.A. acknowledges the support from the German Research Foundation (DFG) under Grant AS 310/9-1. Support from the Ion Beam Center (IBC) at Helmholtz-Zentrum Dresden-Rossendorf (HZDR) is gratefully acknowledged for the proton irradiation.
\end{acknowledgments}


%

\end{document}


\title{Dynamical Reorientation of Spin Multipoles in Silicon Carbide by Transverse Magnetic Fields}
\author{A.~Hern\'{a}ndez-M\'{i}nguez}
\affiliation{Paul-Drude-Institut f\"{u}r Festk\"{o}rperelektronik, Leibniz-Institut im Forschungsverbund Berlin e.V., Hausvogteiplatz 5-7, 10117 Berlin, Germany}
\author{A.~V.~Poshakinskiy}
\affiliation{ICFO-Institut de Ciencies Fotoniques, The Barcelona Institute of Science and Technology, 08860 Castelldefels, Barcelona, Spain}
\author{M.~Hollenbach}
\affiliation{Helmholtz-Zentrum Dresden-Rossendorf, Institute of Ion Beam Physics and Materials Research, Bautzner Landstrasse 400, 01328 Dresden, Germany}
\author{P.~V.~Santos}
\affiliation{Paul-Drude-Institut f\"{u}r Festk\"{o}rperelektronik, Leibniz-Institut im Forschungsverbund Berlin e.V., Hausvogteiplatz 5-7, 10117 Berlin, Germany}
\author{G.~V.~Astakhov}
\affiliation{Helmholtz-Zentrum Dresden-Rossendorf, Institute of Ion Beam Physics and Materials Research, Bautzner Landstrasse 400, 01328 Dresden, Germany}

\begin{abstract}
This Supplemental Material contains information about:
\begin{itemize}
\item Experimental details of the ODMR measurements
\item Husimi spin distributions
\item Kinetic equations for the spin density matrices
\item Rate equations for the spin-level populations
\item Extracting dipole and quadrupole spin polarization from the ODMR signal
\end{itemize}
\end{abstract}

\maketitle

\section{Experimental details of the ODMR measurements}\label{app_ExpDet}

The sample consists of a $10\times10$~mm$^2$ semi-insulating 4H-SiC substrate containing an ensemble of $\VSi$ centers at a depth of 2.5~$\mu$m below the surface. The $\VSi$ centers were created by proton irradiation with an energy of 375~keV and a fluence of $10^{15}$~cm$^{-2}$. The sample was placed on a coplanar wave guide for microwave (MW) field excitation with a nominal power of 30~dBm using a MW signal generator connected to a high-power amplifier. The ODMR experiments were performed at room temperature in a confocal $\mu$-PL setup. The $\VSi$ centers were optically excited by a 780~nm laser beam focused to a spot size of 10~$\mu$m by a $20\times$ objective with numerical aperture NA=0.4. The $\VSi$ PL band centered around 917~nm was collected by the same objective, spectrally separated from the reflected laser beam using an 805~nm dichroic mirror and an 850~nm long-pass filter, and detected by a silicon photodiode. The output signal of the photodiode was locked to the amplitude modulation frequency of the MW signal. The sample and the coplanar waveguide were placed between the poles of an electromagnet applying an external magnetic field $\vBext=(B_x,0,0)$ perpendicular to the $c$-axis of the 4H-SiC substrate. The magnetic field of the MW excitation is perpendicular to both the $c$-axis and $\vBext$. 

\section{Husimi spin distributions}\label{app_Husimi}

To illustrate the quadrupole spin polarization and its transformation under a transverse magnetic field, we show in Fig.~1(b) of the manuscript the distributions of the Husimi quasiprobability densities, $P(\theta,\varphi)_{g,e}$, on the surface of a sphere, for two magnetic field strengths. These distributions are obtained from the 4$\times$4 spin-density matrices $\rho_{g,e}$ of the model described below by calculating the average

\begin{equation}
P(\theta,\varphi)_{g,e} = \bra{3/2}_{\theta,\varphi} \ \rho_{g,e} \ \ket{3/2}_{\theta,\varphi}.
\end{equation}

\noindent Here, $\ket{3/2}_{\theta,\varphi}$ is the coherent spin state with the spin projection $+3/2$ on the direction determined by the polar and azimuthal angles $\theta$ and $\varphi$, that is

\begin{equation}
\ket{3/2}_{\theta,\varphi} = \e^{-\rmi \varphi S_z } \e^{-\rmi \theta S_y } \ket{+3/2}_z.
\end{equation}

\noindent Note that, although $P(\theta,\varphi)$ is always positive, it is not the conventional probability distribution, since the states $\ket{3/2}_{\theta,\varphi}$ with different $\theta$ and $\varphi$ are not orthogonal and, therefore, not mutually exclusive.

\section{Kinetic equations for the spin density matrices}\label{app:KinetEq}

\begin{figure}
\includegraphics[width=.5\linewidth]{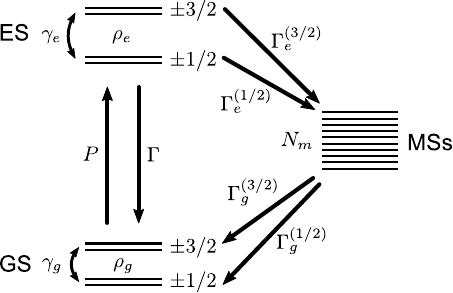}
\caption{Scheme of the electronic levels of the spin center. It includes all transition rates appearing in the kinetic equations~\eqref{eq:sys} .}
\label{figS1}
\end{figure}

The state of the 3/2-spin vacancy is described by the 4$\times$4 density matrices $\rho_g$ and $\rho_e$, corresponding to the GS and ES,  respectively, and the population of the MSs, $N_m$. The spin dynamics is governed by the following set of kinetic equations
\begin{align}\label{eq:sys} \nonumber
\frac{d\rho_g}{dt} &= \frac{\rmi}{\hbar} [\rho_g, H_g] + \Gamma \rho_e - P \rho_g - \gamma_g(\rho_g - {\rm Tr\,}\rho_g/4)\\ \nonumber
&+\frac{N_m}{2} (\Gamma_g^{(1/2)}\mathcal{P}_{1/2}+\Gamma_g^{(3/2)}\mathcal{P}_{3/2}) 
, \nonumber\\ 
\frac{d\rho_e}{dt} &= \frac{\rmi}{\hbar} [\rho_e, H_e] - \Gamma \rho_e- \gamma_e(\rho_e - {\rm Tr\,}\rho_e/4) \nonumber\\ 
& + P \rho_g - \Gamma_e^{(1/2)}\{\mathcal{P}_{1/2}, \rho_e\} - \Gamma_e^{(1/2)}\{\mathcal{P}_{3/2}, \rho_e\} , \nonumber \\
\frac{dN_m}{dt} &=
\Gamma_e^{(1/2)} {\rm Tr\,}(\mathcal{P}_{1/2} \rho_e) + \Gamma_e^{(3/2)} {\rm Tr\,}(\mathcal{P}_{3/2} \rho_e)  \nonumber \\
&- (\Gamma_g^{(1/2)}+\Gamma_g^{(3/2)})N_m,
\end{align}
\noindent together with the normalization condition ${\rm Tr\,} \rho_g+ {\rm Tr\,} \rho_e + N_m = 1$. Here, $H_{g(e)}$ are the Hamiltonians of the GS and ES given by Eq.~(1) in the manuscript, $P$ and $\Gamma$ the optical pump and recombination rates, 
\begin{align}
&\Gamma_{g}^{1/2(3/2)} = \Gamma_{e} (1 \pm \eta_g), \\
&\Gamma_{e}^{1/2(3/2)} = \Gamma_{e} (1 \pm \eta_e)
\end{align}
the spin-dependent relaxation rates between the MSs and the $\pm 1/2$ ($\pm 3/2$) states in ES and GS, and $\gamma_{g,e}$ the spin relaxation rates within GS and ES, which are assumed isotropic. We also used the notation $\{A,B\} = (AB+BA)/2$, $[A,B] = AB-BA$, $\mathcal{P}_{1/2}=\ket{+1/2}_z\bra{+1/2}_z+\ket{-1/2}_z\bra{-1/2}_z$ and $\mathcal{P}_{3/2}=\ket{+3/2}_z\bra{+3/2}_z+\ket{-3/2}_z\bra{-3/2}_z$.

\section{Rate equations for the spin-level populations}\label{app_RateEq}

When the fine structure splittings of the spin centers are much larger than the relaxation or pump rates, we can neglect the off-diagonal components of the density matrices $\rho_{g,e}$. Then, the spin state is described by the population of the four eigestates in the GS and ES, $\bm f_g=(f_{g1},f_{g2},f_{g3},f_{g4})$ and $\bm f_e=(f_{e1},f_{e2},f_{e3},f_{e4})$, which are ordered according to their energies in descending order. We introduce the population variations 
$\delta f_{g(e)}^{(i)}=f_{g(e)}^{(i)} - \sum_i f_{g(e)}^{(i)}/4$, which satisfy the following set of equations in the steady state
\begin{align}\label{eq:df}
& \eta_g \Gamma_e N_e \hat{ \bm T}_{g0} \bm d_0 - (P+\gamma_g)\bm{\delta f_g} + \Gamma \hat{\bm T}_{ge} \bm{\delta f_e} =0 \,,\\
  -&\eta_e \Gamma_e N_e \hat{\bm T}_{e0} \bm d_0 + P\hat{\bm T}_{eg}\bm{\delta f_g} - (\Gamma+\Gamma_e+\gamma_e)  \bm{\delta f_e} =0\,, \nonumber
\end{align}
\noindent where $N_e = {\rm Tr\,} \rho_e $ is the population of the excited state, $\bm d_0 = (-1,-1,1,1)/2$ the population variation corresponding to the spin quadrupole in zero magnetic field, and 
 $\hat T^{ij}_{g0} = |U^{ij}_{g0}|^2$, $\hat T^{ij}_{e0} = |U^{ij}_{e0}|^2$, $\hat T^{ij}_{eg} =\hat T^{ji}_{ge} = |U^{ij}_{eg}|^2$. Here, $U^{ij}_{g0}$ ($U^{ij}_{e0}$) is the unitary matrix relating the eigenstates in GS(ES) under $\Bext\neq0$ to the corresponding eigenstates in $\Bext=0$, while $U^{ij}_{eg}$ relates the eigenstates in the ES to those in the GS under $\Bext\neq0$. 

The solution of Eq.~\eqref{eq:df} reads
\begin{align}
\bm f_e =&\frac{\Gamma_e N_e}{\Gamma+\Gamma_e+\gamma_e} \left( 1 - \frac{P\Gamma}{(P+\gamma_g)(\Gamma+\Gamma_e+\gamma_e)}\hat{\bm T}_{eg}\hat{ \bm T}_{ge}\right)^{-1} 
\left( \eta_g \frac{P}{P+\gamma_g} \hat{\bm T}_{eg} \hat{\bm T}_{g0} - \eta_e \hat{\bm T}_{e0}\right) \bm d_0 \,,
\end{align}
\noindent and the PL intensity variation is given by
\begin{equation}
\Delta\text{PL} =  -\eta_e \bm{d}_0 \cdot \hat{\bm T}_{0e} \bm{\delta f}_e \,.
\end{equation}

Application of a MW field with frequency matching the energy difference between a pair of spin sublevels $i$ and $j$ inside GS or ES leads to a change in the population of the spin eigenstates
\begin{equation}\label{eq:df2}
(\dot f_{g,e})_k \propto |M_{ij}|^2 [(\delta f_{g,e})_j - (\delta f_{g,e})_i](\delta_{i,k}- \delta_{j,k}) \,,
\end{equation}
\noindent where $M_{ij}$ is the matrix element of the spin transition. Such variation of population leads to a corresponding change of the ES quadrupole, which determines the intensity of the corresponding ODMR lines for ES and GS:

\begin{align}\label{eq:ODMRe}
&\Delta\text{PL} _{e} =  
-\eta_e \bm{d}_0 \cdot \hat{\bm T}_{0e} \left[ 1 - \frac{P\Gamma}{(P+\gamma_g)(\Gamma+\Gamma_e+\gamma_e)}\hat{\bm T}_{eg}\hat{ \bm T}_{ge}\right]^{-1} \bm{\dot{ f}}_{e} \,,\\
&\Delta\text{PL} _{g} = 
-\eta_e \bm{d}_0 \cdot \hat{\bm T}_{0e}\hat{\bm T}_{eg} \left[ 1 - \frac{P\Gamma}{(P+\gamma_g)(\Gamma+\Gamma_e+\gamma_e)}\hat{\bm T}_{eg}\hat{ \bm T}_{ge}\right]^{-1} \bm{\dot{ f}}_{g}
\end{align}

\subsection*{Small magnetic fields}

The application of a small magnetic field $g\mu_BB_x \ll D^{(e)}$ affects the GS eigenstates, while the ES eigenstates remain almost unperturbed. In this situation, the quadrupole momentum of the ES reads

\begin{align}
&\bm d_0 \cdot  \bm{\delta f}_e= \frac{\eta_g}{\Gamma+\Gamma_e+\gamma_e} \left( 1 - \frac{xP\Gamma}{(P+\gamma_g)(\Gamma+\Gamma_e+\gamma_e)} \right)^{-1} 
\left(\frac{xP}{P+\gamma_g}  - \frac{\eta_e}{\eta_g} \right), \  \
\end{align}

\noindent with 
\begin{align}
   x =  \frac{1}{4} \left( 1+ \frac3{1+b_g^2+b_g^4}\right)
\end{align}
being the eigenvalue of the matrix $\hat{\bm T}_{0g} \hat{\bm T}_{g0}$, that satisfies the equation $\hat{\bm T}_{0g} \hat{\bm T}_{g0} \bm{d}_0 = x \bm{d}_0$ and  describes the loss of the spin quadrupole upon its transition from ES to GS and back. We also introduced  $b_g=g\mu_BB_x/D^{(g)}$. 

In the particular case of $\gamma_e,\gamma_g, \Gamma_e \ll P, \Gamma$, the ODMR signal for ES assumes the simple form 

\begin{align}
\Delta\text{PL} _e \sim \frac{\Gamma_e N_e}{(1-x)\Gamma} \eta_e \eta_g \left( x - \frac{\eta_e}{\eta_g} \right) \,.
\end{align}

\noindent Note that, if $ 1/4 <\eta_e/\eta_g<1$, the ODMR signal for ES changes sign when $x = \eta_e/\eta_g$.

The sign of the ODMR signal for all spin transitions within the GS is determined by
 
\begin{align}
\Delta\text{PL}_g \sim \eta_e\eta_g\left(1-\frac{\eta_e}{\eta_g}\right).
\end{align}

\subsection*{Large magnetic fields}

At large magnetic fields, $g\mu_BB_x \gg D^{(g)}$, the GS zero-field splitting can be neglected, and the spin in the GS is quantized along the $x$ axis. In this case, the ODMR signal intensities for the three observed GS spin transitions are given by

\begin{align}
& \Delta\text{PL}_g^{\pm 1/2 \leftrightarrow \pm 3/2} \sim  \frac{\eta_e \eta_g }4  \, (1+b_e^2) \left(1 -\frac{2-b_e^2+2b_e^4}{2(1+b_e^2+b_e^4)} \frac{\eta_e}{\eta_g} \right)  \,,\nonumber\\
& \Delta\text{PL}_g^{- 1/2 \leftrightarrow +1/2} \sim -\frac{5\eta_e^2}4 \,  \frac{b_e^2(1+b_e^2)}{1+b_e^2+b_e^4} \,,
\end{align}

\noindent with $b_e=g\mu_BB_x/D^{(e)}$. While the ODMR signal for the $-1/2 \leftrightarrow +1/2$ transition is always negative at large magnetic fields, the ODMR signal of the $\pm 1/2 \leftrightarrow \pm 3/2$ transitions is positive if $\eta_e/\eta_g<1$, as observed in our experiment where $\eta_e/\eta_g\approx0.7$, negative if $\eta_e/\eta_g>2$, and would change sign twice at certain magnetic fields $b^*$ and $1/b^*$ if $1<\eta_e/\eta_g<2$. 

\section{Extracting dipole and quadrupole spin polarization from the ODMR signal}\label{app_SpinExt}

From the experimentally measured intensities of the ODMR signal (peak areas), we can extract the population difference between the various pairs of spin sublevels.
In the experimental spectra, three GS spin transitions between the $1 \leftrightarrow 2$, $2 \leftrightarrow 3$, and $3 \leftrightarrow 4$ eigenstates are observed (as in the previous section, the spin eigenstates are numbered according to their energy in decreasing order). With the matrix elements and the brightness of all the levels, we extract the population variation for all four GS spin sublevels, $\delta f_g^{(i)}$, from the intensities (resonance peak areas) of three GS transitions and taking into account the constraint $\sum_i \delta f_g^{(i)}=0$.
A similar approach is used to get the ES population variations, $\delta f_e^{(i)}$. Since in this case only the ES spin transitions between the $1 \leftrightarrow 4$ and $2 \leftrightarrow 3$ eigenstates are observed, the remaining transitions are prescribed with zero intensity.  

Once $\delta f_g^{(i)}$ and $\delta f_e^{(i)}$ are obtained, the quadrupole and dipole spin polarizations are given by
\begin{align}
\langle \delta S_z^2\rangle =& \frac{(2+b)(\delta f_2-\delta f_4)}{4\sqrt{1+b+b^2}} +\frac{(2-b)(\delta f_1-\delta f_3)}{4\sqrt{1-b+b^2}} ,\nonumber\\
\langle  S_x\rangle = &\frac{(1+2b)(\delta f_2-\delta f_4)}{4\sqrt{1+b+b^2}} +\frac{(1-2b)(\delta f_3-\delta f_1)}{4\sqrt{1-b+b^2}}  \nonumber \\
&+\frac12(\delta f_1+\delta f_3-\delta f_2-\delta f_4),
\end{align}
where $b=g\mu_BB_x/D^{(g,e)}$ for GS and ES, respectively.